\documentclass{IEEEtran}
\usepackage{amssymb}
\usepackage{amsfonts}
\usepackage{amsmath}
\usepackage{algorithm}
\usepackage{algorithmic}
\usepackage{graphicx,cite,multicol,caption,booktabs,balance}
\usepackage{color}

\begin{document}
\title{\huge UAV-Enabled Uplink Non-Orthogonal Multiple Access System: Joint Deployment and Power Control}

\author{Jinhui Lu, Yuntian Wang, Tingting Liu, \IEEEmembership{Member,~IEEE,} Zhihong Zhuang, Xiaobo Zhou,  Feng Shu, \IEEEmembership{Member,~IEEE,} and Zhu Han, \IEEEmembership{Fellow,~IEEE,}

\thanks{This work was supported in part by the National Natural Science Foundation
of China under Grants 61801453 (Corresponding authors: Zhihong Zhuang, Tingting Liu, and Feng Shu).}
\thanks{~Jinhui Lu, Yuntian Wang, Xiaobo Zhou, Zhihong Zhuang and Feng Shu are with School of Electronic and Optical Engineering, Nanjing University of Science and Technology, Nanjing, 210094, China (e-mail: lujinhui\_{}njust\_{}1962@163.com; wytpzaaa@njust.edu.cn; zxb@njust.edu.cn; nustcn@163.com; shufeng@njust.edu.cn).}
\thanks{~Tingting Liu is with School of Information and Communication Engineering, Nanjing Institute of Technology, Nanjing, 211167, China (e-mail: liutt@njit.edu.cn).}
\thanks{~Zhu Han is with the Department of Electrical and Computer Engineering,
University of Houston, Houston, TX 77004, USA (e-mail: zhan2@uh.edu).}}

\maketitle

\begin{abstract}
In order to overcome the inherent latency in multi-user unmanned aerial vehicle (UAV) networks with orthogonal multiple access (OMA). In this paper, we investigate the UAV enabled uplink non-orthogonal multiple access (NOMA) network, where a UAV is deployed to collect the messages transmitted by ground users. In order to maximize the sum rate of all users and to meet the quality of service (QoS) requirement, we formulate an optimization problem, in which the UAV deployment position and the power control are jointly optimized. This problem is non-convex and some variables are binary, and thus it is a typical NP hard problem. In this paper, an iterative algorithm is proposed with the assistance of successive convex approximate (SCA) technique and the penalty function method. In order to reduce the high computational complexity of the iterative algorithm, a low complexity approximation algorithm is then proposed, which can achieve a similar performance compared to the iterative algorithm. Compared with OMA scheme and conventional NOMA scheme, numerical results show that our proposed algorithms can efficiently improve the sum rate.
\end{abstract}

\begin{IEEEkeywords}
UAV networks, Uplink NOMA, power control, UAV deployment.
\end{IEEEkeywords}

\section{Introduction}
Non-orthogonal multiple access (NOMA) has received substantial attention because of its superiorities in low latency, spectrum efficiency, and connectivity of users compared with orthogonal multiple access (OMA) \cite{7263349,6692652,7973036}. Different from OMA schemes such as frequency division multiple access (FDMA) and time division multiple access (TDMA), users with NOMA share the same time, frequency and code resources simultaneously but have difference in power domain, which inevitably introduces inter-user interference. As such, superposition coding (SC) is used at transmitters, and successive interference cancellation (SIC) is deployed at receivers \cite{7263349,6692652,7973036,5374051,5173127}. Downlink NOMA networks have been studied in \cite{6868214} and \cite{7069272}. It has been proven that NOMA can achieve better outage performance than OMA, if users' rate and power allocation are carefully designed. Then, the authors in \cite{7390209} and \cite{6933459} discussed uplink NOMA networks and demonstrated that uplink NOMA can improve spectrum efficiency and fairness index compared with OMA. Afterwards, a novel cooperative NOMA scheme was proposed in \cite{7117391}, in which the users with stronger channel gains can be used as relays to improve the performance of the users with poorer channel gains by making full use of SIC. Moreover, to reduce the complexity of SIC, NOMA schemes with user pairing were raised in \cite{7273963} and \cite{8436709}, where every two users were paired to carry out NOMA while OMA was executed for different user pairs.

In recent years, the research enthusiasm for unmanned aerial vehicles (UAVs) networks has risen sharply due to its advantages of high mobility, flexible deployment, low cost and high probability of line-of-sight (LoS) link \cite{7470933,8254949,8247211,8119562,8643815,8254971,8432499,7888557,8453022,7572068}. For example, in \cite{8254949,8247211,8119562}, the UAVs were employed as mobile base stations or mobile data collectors to enhance the coverage and communication quality. Then, the issues of physical layer security have been researched for UAV networks in \cite{8643815,8254971,8432499}. Besides, in order to prolong the network lifetime, the energy-efficient UAV networks were studied in \cite{8119562} and \cite{7888557}. Furthermore, the UAVs were utilized as mobile relays to provide connections between users when there is no reliable direct link in \cite{8453022} and \cite{7572068}.

However, according to \cite{8254949,8304088} and \cite{8438896}, there exists a fundamental tradeoff between delay and throughput in multi-user UAV networks with OMA. In order to reduce the access latency and to further improve the communication quality of the UAV networks, it is natural to consider applying NOMA to the UAV networks, namely UAV-enabled NOMA networks. There have been a few works related to the UAV-enabled downlink NOMA networks. In \cite{8647352}, the authors proposed an algorithm to maximize the minimum average rate through jointly optimizing the UAV trajectory and power allocation. In addition, a UAV-enabled NOMA network with user pairing has been studied in \cite{8672190} to maximize the minimum throughput. Furthermore, to maximize the sum rate of the ground users, the authors in \cite{8663350} studied the UAV placement location and power allocation for the UAV-enabled NOMA network. However, the extension from downlink NOMA to uplink NOMA is not trivial because the decoding order of SIC in uplink NOMA is completely opposite to that of downlink NOMA. Besides, the messages transmitted by different users in uplink NOMA experience different channel gains when arriving the base station, while the messages intended for different users in downlink NOMA experience the same channel gain when reaching the target user. Note that the aforementioned literatures only considered the downlink scenarios, so they can-not be directly applied to the uplink scenarios such as data collection in the upcoming Internet of Things (IoT).

Motivated by the above reason, a UAV-enabled uplink NOMA network is considered in this paper, where a UAV is deployed to collect the messages transmitted by the ground users. Our target is to maximize the sum rate of all users through jointly designing the UAV trajectory and the power control. The consequent problem is a mixed integer non-convex optimization problem which is NP hard to solve straightforwardly. Thus, we propose an iterative algorithm to solve it with the help of the successive convex approximation (SCA) technique and the penalty function method. The main contributions of this work are summarized as below.
\begin{itemize}
\item First, we consider a UAV-enabled NOMA system in scenario of uplink. The concerned scenario is different from the UAV-enabled downlink NOMA scenarios. Numerical results demonstrate that the conventional uplink NOMA network which has fixed base stations can be greatly improved by introducing UAVs to the system from the perspective of the sum rate of users.
\item We prove that the UAVs need to stay stationary at a certain point in our system to obtain the optimal system performance, which transforms the problem from optimizing the UAV trajectory to searching for the optimal UAV deployment position. Besides, the computational complexity of the problem can be significantly reduced.
\item Although the computational complexity of the problem has been significantly reduced, it is time-consuming to obtain the optimal solution and it is difficult to set initial feasible points. Therefore, we propose a low complexity approximation algorithm when the required user rate is not very large.
\item Numerical results show that the proposed approximation algorithm can greatly enhance the sum rate performance compared to NOMA schemes with fixed base stations, and can achieve less than 4\% performance loss compared to the proposed iterative algorithm. The most important is that the approximation algorithm only takes a few seconds.
\end{itemize}

The rest of the paper is organized as follows. In Section \ref{System Model and Problem Formulation}, we present the system model for the UAV-enabled uplink NOMA network and formulate the optimization problem. Then, an iterative algorithm is proposed to solve the optimization problem by the SCA technique and the penalty function method in Section \ref{Iterative Algorithm For Problem (P1)}. In Section \ref{Initialization for algorithm 1}, an initialization scheme for our proposed iterative algorithm and a low complexity approximation algorithm for the optimization problem are presented. Afterwards, in Section \ref{Numerical Results}, numerical results are provided to demonstrate the performance improvement of our proposed algorithms compared with the benchmarks. Finally, the paper is concluded in Section \ref{Conclusions}.

\section{System Model and Problem Formulation}
\label{System Model and Problem Formulation}
\begin{figure}[tp]
  \centering
  \includegraphics[width=0.45\textwidth]{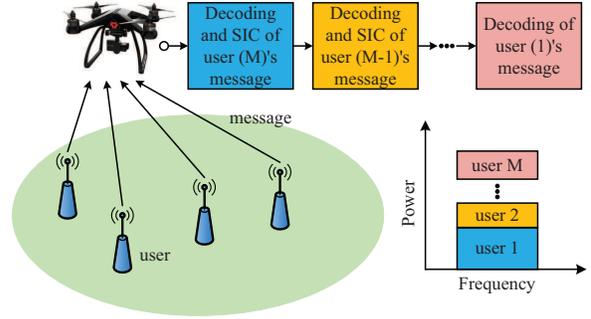}
  \captionsetup{font=footnotesize}
  \caption{UAV-Enabled Uplink NOMA System.}
  \label{sys-model}
\end{figure}

\subsection{System Model}\

As shown in Fig. \ref{sys-model}, we consider a UAV-enabled uplink NOMA system, where $M$ users are located on the ground and a UAV is employed as a mobile base station to serve the $M$ users periodically with NOMA. The flight period and the maximum flight speed of the UAV are denoted as $T$ in second (s) and $V_\mathrm{max}$ in meter/second (m/s), respectively. Besides, it is assumed that the UAV flies at a fixed altitude $H$ in meters (m). Without loss of generality, we consider a 3-D Cartesian coordinate system, where the horizontal coordinates of the $i$-th user and the UAV at time $t$ are denoted by $\mathbf{q}_i=[x_i,y_i]^T, i\in\mathcal{I}=\{1,2,\ldots, M\}$ and $\mathbf{q}(t)=[x(t),y(t)]^T, 0\leq t\leq T$, respectively. For ease of exposition, we evenly divide the flight period $T$ into $N$ sufficiently small time slots (the duration of each time slot is given by $\delta_t=\frac{T}{N}$), so that the UAV can be considered to be static in each time slot. As a result, the trajectory of the UAV can be approximated as $\mathbf{q}[n]=[x[n],y[n]]^T, n\in\mathcal{N}=\{1,2,\ldots,N\}$. Based on the above assumptions, the UAV mobility constraints can be expressed as
\begin{subequations}\label{eq1}
\begin{equation}\label{1a}
\mathbf{q}[1]=\mathbf{q}[N],
\end{equation}
\begin{equation}\label{1b}
\|\mathbf{q}[n+1]-\mathbf{q}[n]\|^2\leq L^2, n=1,2,\ldots, N-1,
\end{equation}
\end{subequations}
where $L=V_\mathrm{max}\delta_t$ is the maximum UAV flight distance in each time slot. (\ref{1a}) ensures that the UAV has to return to the initial point after finishing a cycle of flight, aiming to serve the ground users periodically. (\ref{1b}) indicates the speed of the UAV can-not exceed the maximum flight speed.

For simplicity, we assume that all the nodes in the system are equipped with a single antenna and the communication links from the users to the UAV are dominated by Line-of-Sight (LoS) links. Besides, it is assumed that the Doppler effect is perfectly compensated at the UAV. As such, the channel gain from the $i$-th user to the UAV in time slot $n$ follows free space path loss, which can be modeled as
\begin{equation}\label{eq2}
h_i[n]=\sqrt{\beta_0d_i^{-2}[n]}=\sqrt{\frac{\beta_0}{H^2+\|\mathbf{q}[n]-\mathbf{q}_i\|^2}}, \forall i,n,
\end{equation}
where $\beta_0$ denotes the channel gain at the reference distance, i.e., $d_0$=1 m when the transmission power is equal to $1$ W and
\begin{equation}\label{eq3}
d_i[n]=H^2+\|\mathbf{q}[n]-\mathbf{q}_i\|^2,\forall i,n,
\end{equation}
is the distance from the $i$-th user to the UAV in time slot $n$. According to the NOMA principle, in time slot $n$, the UAV receives the superposition message
\begin{equation}\label{eq4}
y[n]=\sum_{i=1}^{M} \sqrt{P_i[n]}h_i[n]x_i[n]+n_u,\forall n,
\end{equation}
where $x_i[n]$ denotes the message that the $i$-th user sends to the UAV in time slot $n$ with the transmission power $P_i[n]$, and $n_u$ denotes the zero-mean additive white Gaussian noise (AWGN) with the variance $\sigma^2$ at the UAV. In order to reduce the inter-user interference, the transmission power should satisfy the maximum total transmission power constraint \cite{7542118}, which can be expressed as
\begin{subequations}\label{eq5}
\begin{equation}\label{5a}
P_i[n]\geq0, \forall i,n,
\end{equation}
\begin{equation}\label{5b}
\sum_{i=1}^{M}P_i[n]\leq P_\mathrm{max}, \forall n,
\end{equation}
\end{subequations}
where $P_\mathrm{max}$ is the maximum total transmission power of all users.

The UAV employs SIC to decode the messages from different users. Specifically, the messages from the users with poorer channel gains are treated as interference for the users with stronger channel gains. While the messages from the stronger users have been subtracted from the received message when decoding the messages from the poorer suers. For ease of exposition, we introduce variables $\alpha_{ij}[n]\in\{0,1\}, \forall i,j\in\mathcal{I}, n\in\mathcal{N}$ to denote the SIC decoding order, where $\alpha_{ij}[n]=1$ implies that the $j$-th user has poorer channel gain, and the $j$-th user's message is treated as interference when decoding the $i$-th user's message in time slot $n$; otherwise, $\alpha_{ij}[n]=0$. As a result, $\alpha_{ij}[n]$ are defined as follow:
\begin{equation}\label{eq6}
\alpha_{ij}[n]=\left\{
\begin{aligned}
&0,&d_i[n]>d_j[n],  \\
&1,&d_i[n]<d_j[n],  \\
&0~\mathrm{or}~1,&d_i[n]=d_j[n].
\end{aligned}
\right.
\end{equation}
(\ref{eq6}) exploits the distance to represent the relationship between the channel gains equivalently. However, variables $\alpha_{ij}[n]$ are defined by the boolean operation, which is not tractable. Therefore, we rewrite it as (\ref{eq7}).
\begin{subequations}\label{eq7}
\begin{equation}\label{eq7a}
\alpha_{ij}[n]\in\{0,1\}, \forall i\neq j,n,
\end{equation}
\begin{equation}\label{eq7b}
\alpha_{ii}[n]=0, \forall i,n,
\end{equation}
\begin{equation}\label{eq7c}
\alpha_{ij}[n]+\alpha_{ji}[n]=1, \forall i\neq j,n
\end{equation}
\begin{equation}\label{eq7d}
\alpha_{ij}[n](H^2+\|\mathbf{q}[n]-\mathbf{q}_i\|^2)\leq (H^2+\|\mathbf{q}[n]-\mathbf{q}_j\|^2), \forall i\neq j,n.
\end{equation}
\end{subequations}
(\ref{eq7a}) represents that for two different users, $\alpha_{ij}[n]$ is either 1 or 0. (\ref{eq7b}) indicates that the UAV should not treat the message from the $i$-th user as interference when decoding the $i$-th user's message. (\ref{eq7c}) means that for two different users, if one is considered as the stronger user, the other has to be considered as the poorer user. (\ref{eq7a}), (\ref{eq7c}) and (\ref{eq7d}) ensure that if $d_i[n]>d_j[n]$, then $\alpha_{ij}[n]=0$; otherwise $\alpha_{ij}[n]=1$.

We assume that the available bandwidth is normalized. Then the achievable rate from the $i$-th user to the UAV in time slot $n$ in bits/second/Hertz (bps/Hz) is given by
\begin{equation}\label{eq8}
R_i[n]=\mathrm{log}_2\left(1+\frac{P_i[n]\tilde{h}_i[n]}{1+\sum_{j=1,j\neq i}^{M} \alpha_{ij}[n]\tilde{h}_j[n]P_j[n]}\right), \forall i,n,
\end{equation}
where $\tilde{h}_i[n]=\frac{h_i^2[n]}{\sigma^2}=\frac{\gamma_0}{H^2+\|\mathbf{q}[n]-\mathbf{q}_i\|^2}$, $\gamma_0=\frac{\beta_0}{\sigma^2}$. The numerator and denominator of the fraction inside logarithm operation (\ref{eq8}) denote the power of desired signal and the power of interference and noise when decoding the $i$-th user's message in time slot $n$, respectively. The average achievable rate from the $i$-th user to the UAV in bps/Hz over $N$ time slots is given by
\begin{equation}\label{eq9}
R_i=\frac{1}{N}\sum_{n=1}^{N}R_i[n], \forall i.
\end{equation}

\subsection{Problem Formulation}

Our objective is to maximize the summation of the average rates over $M$ users with the users quality of service (QoS) constraint through jointly optimizing the UAV trajectory and the power control. The optimization problem can be formulated as
\begin{subequations}\label{P1}
\begin{equation}\label{10a}
\mathrm{(P1):}\max_{\mathbf{Q}, \mathbf{P}, \mathbf{A}}~\sum_{i=1}^{M}\frac{1}{N}\sum_{n=1}^{N}R_i[n],
\end{equation}
\begin{equation}\label{10b}
\text{s.t.}~~R_i[n]\geq r^*,\forall i,n,\\
\end{equation}
\begin{equation}\label{10c}
(\ref{eq1}), (\ref{eq5}), (\ref{eq7}),
\end{equation}
\end{subequations}
where $\mathbf{Q}=\{\mathbf{q}[n], \forall n\}$ is the UAV trajectory, $\mathbf{P}=\{P_i[n], \forall i,n\}$ is the power control, $\mathbf{A}=\{\alpha_{ij}[n], \forall i,j,n\}$ is the decoding order, and $r^*$ denotes the instantaneous rate threshold for all users. Note that in \cite{8629316}, the authors used the average rate to guarantee QoS, causing some users can-not access to the UAV in some time slots. Instead, constraint (\ref{10b}) ensures that all the users can communicate with the UAV in each time slot, which consequently achieves a low access latency.

Problem (P1) is difficult to solve due to the following two reasons. First, constraints (\ref{eq7d}) and (\ref{10b}) are non-convex with respect to $\mathbf{Q}$, $\mathbf{P}$ and $\mathbf{A}$. Second, the variables $\alpha_{ij}$ are binary. As such, problem (P1) is a mixed integer non-convex problem, which is NP hard. In Section \uppercase\expandafter{\romannumeral3}, we will propose an iterative algorithm to find a solution.

\section{Iterative Algorithm For Problem (P1)}
\label{Iterative Algorithm For Problem (P1)}

In this section, we propose an iterative algorithm to solve problem (P1). Specifically, we first prove that problem (P1) can be simplified by removing the time variable $n$. Then we deal with the binary constraints based on the penalty function method, and transform the non-convex constraints into convex ones based on SCA technique. As a result, the NP hard problem (P1) can be transformed into a tractable convex optimization problem, which can be solved by CVX \cite{cvx}.

\subsection{Problem Simplification}

Note that in any time slot $n$, the sum rate of all users is independent of the decoding order \cite{6933459}, then we can obtain
\begin{equation}\label{eq11}
\begin{aligned}
\sum_{i=1}^{M}R_i[n]&=\sum_{i=1}^{M}\log_2\left(1+\frac{P_i[n]\tilde{h}_i[n]}{1+\sum_{j=i+1}^{M} \tilde{h}_j[n]P_j[n]}\right)\\
&=\sum_{i=1}^{M}\log_2\left(\frac{1+\sum_{j=i}^{M} \tilde{h}_j[n]P_j[n]}{1+\sum_{j=i+1}^{M} \tilde{h}_j[n]P_j[n]}\right)\\
&=\log_2\left(1+\sum_{i=1}^{M}P_i[n]\tilde{h}_i[n]\right),\forall n.
\end{aligned}
\end{equation}
Base on (\ref{eq11}), problem (P1) is equivalent to problem (P2) as follows.
\begin{subequations}\label{P2}
\begin{equation}\label{12a}
\mathrm{(P2):}\max_{\mathbf{Q}, \mathbf{P}, \mathbf{A}}~\frac{1}{N}\sum_{n=1}^{N}\log_2\left(1+\sum_{i=1}^{M}P_i[n]\tilde{h}_i[n]\right),
\end{equation}
\begin{equation}\label{12b}
\text{s.t.}~(\ref{eq1}), (\ref{eq5}), (\ref{eq7}), (\ref{10b}).
\end{equation}
\end{subequations}
For problem (P2), we have the following proposition.
\newtheorem{proposition}{\textbf{Proposition}}
\begin{proposition}
To maximize the summation of the average rates over $M$ users, the UAV should stay stationary at a certain point.
\end{proposition}
\begin{IEEEproof}
We proof Proposition 1 by contradiction.

First, $\forall n\in\mathcal{N}$, considering the following problem.
\begin{subequations}\label{P3}
\begin{equation}\label{13a}
\mathrm{(P3):}\max_{\mathbf{q}[n], \mathbf{P}[n], \mathbf{A}[n]}\log_2\left(1+\sum_{i=1}^{M}P_i[n]\tilde{h}_i[n]\right),
\end{equation}
\begin{equation}\label{13b}
\text{s.t.}~~(\ref{eq5}), (\ref{eq7}), (\ref{10b}),
\end{equation}
\end{subequations}
where $\mathbf{q}[n]$, $\mathbf{P}[n]$ and $\mathbf{A}[n]$ are the position of the UAV, the power control and the decoding order in time slot $n$, respectively. Then on the condition that problem (P3) is feasible (which is a necessary condition for problem (P2) to be feasible), there certainly exist a $\{\mathbf{q}^*[n],\mathbf{P}^*[n],\mathbf{A}^*[n]\}$ satisfying the constraints (\ref{eq5}), (\ref{eq7}), (\ref{10b}) and maximizing the objective function simultaneously, where $\mathbf{q}^*[n]=[x^*[n],y^*[n]]^T$, $\mathbf{P}^*[n]=\{P_i^*[n],\forall i\}$ and $\mathbf{A}^*[n]=\{\alpha_{ij}^*[n],\forall i,j\}$ are the optimal deployment position of the UAV, the optimal power control and the optimal decoding order in time slot $n$, respectively. Note that the unique distinction between the above $N$ optimization problems is the difference in time slot $n$. For this reason, we can obtain that $\{\mathbf{q}^*[1],\mathbf{P}^*[1],\mathbf{A}^*[1]\}=\{\mathbf{q}^*[2],\mathbf{P}^*[2],\mathbf{A}^*[2]\}=\ldots=\{\mathbf{q}^*[N],\mathbf{P}^*[N],\mathbf{A}^*[N]\}\triangleq \{\mathbf{q}^*,\mathbf{p}^*,\mathbf{a}^*\}$. Obviously, $\{\mathbf{Q}^*,\mathbf{P}^*,\mathbf{A}^*\}=\{\mathbf{q}_N^*,\mathbf{p}_N^*,\mathbf{a}_N^*\}$ satisfies constraint (\ref{eq1}) because the UAV is static, where $\mathbf{q}_N^*$ is the matrix with the same size of $\mathbf{Q}$ obtained by replicating $\mathbf{q}^*$, $\mathbf{p}_N^*$ is the matrix with the same size of $\mathbf{P}$ obtained by replicating $\mathbf{p}^*$, and $\mathbf{a}_N^*$ is the matrix with the same size of $\mathbf{A}$ obtained by replicating $\mathbf{a}^*$.  As a result, $\{\mathbf{Q}^*,\mathbf{P}^*,\mathbf{A}^*\}=\{\mathbf{q}_N^*,\mathbf{p}_N^*,\mathbf{a}_N^*\}$ is a feasible solution to problem (P2).

\newcounter{mytempieqncnt}
\begin{figure*}[hbt]
\normalsize
\setcounter{mytempieqncnt}{\value{equation}}
\setcounter{equation}{13}
\begin{equation}\label{eq14}
\begin{aligned}
f(\mathbf{Q},\mathbf{P},\mathbf{A})&=\frac{1}{N}\sum_{n=1}^{N}f_n(\mathbf{q}[n],\mathbf{P}[n],\mathbf{A}[n])\\
&=\frac{1}{N}\left[f_{n^{\prime}}(\mathbf{q}[n^{\prime}],\mathbf{P}[n^{\prime}],\mathbf{A}[n^{\prime}])+\sum_{n=1,n\neq n^{\prime}}^{N}f_n(\mathbf{q}^*,\mathbf{p}^*,\mathbf{a}^*)\right]\\
&\overset{(a)}{\leq} \frac{1}{N}\left[f_{n^{\prime}}(\mathbf{q}^*,\mathbf{p}^*,\mathbf{a}^*)+\sum_{n=1,n\neq n'}^{N}f_n(\mathbf{q}^*,\mathbf{p}^*,\mathbf{a}^*)\right]\\
&=\frac{1}{N}\sum_{n=1}^{N}f_n(\mathbf{q}^*,\mathbf{p}^*,\mathbf{a}^*)\\
&=f_n(\mathbf{q}^*,\mathbf{p}^*,\mathbf{a}^*).
\end{aligned}
\end{equation}
\hrulefill
\setcounter{equation}{14}
\end{figure*}

Assume that there exists some time slot $n^{\prime}$ making $\mathbf{q}[n^{\prime}]\neq \mathbf{q}^*$, $\mathbf{P}[n^{\prime}]$ and $\mathbf{A}[n^{\prime}]$ are the optimal power control and the optimal decoding order in time slot $n^{\prime}$ corresponding to $\mathbf{q}[n^{\prime}]$. Besides, $\forall n\in\mathcal{N}-\{n^{\prime}\}$, we assume $\{\mathbf{q}[n],\mathbf{P}[n],\mathbf{A}[n]\}=\{\mathbf{q}^*,\mathbf{p}^*,\mathbf{a}^*\}$. Denote $f(\mathbf{Q},\mathbf{P},\mathbf{A})=\frac{1}{N}\sum_{n=1}^{N}\log_2(1+\sum_{i=1}^{M}P_i[n]\tilde{h}_i[n])$, $f_n(\mathbf{q}[n],\mathbf{P}[n],\mathbf{A}[n])=\log_2(1+\sum_{i=1}^{M}P_i[n]\tilde{h}_i[n]),\forall n$, then, $f=\frac{1}{N}\sum_{i=1}^{N}f_n$.

As a consequence, we have the inequality (\ref{eq14}) at the top of the next page, where (a) holds since $\{\mathbf{q}^*,\mathbf{p}^*,\mathbf{a}^*\}$ is the optimal solution in time slot $n'$. The inequality (\ref{eq14}) indicates that the sum of the average rates over $M$ users will decrease or remain unchanged if the UAV does not stay stationary at the point $\mathbf{q}^*$. This completes the proof.
\end{IEEEproof}

\newtheorem{remark}{\textbf{Remark}}
\begin{remark}
If there are multiple optimal positions, the UAV should stay stationary at one of the optimal positions. Otherwise, for sufficiently small $\delta_t$, the UAV can-not fly from one optimal position to another optimal position in one time slot.
\end{remark}

According to Proposition 1, optimizing the UAV trajectory is equivalent to searching for the optimal UAV deployment position, i.e., we can remove the time variable $n$ in the original problem (P2). As a sequence, problem (P2) is equivalent to problem (P4).
\begin{subequations}\label{15}
\begin{equation}\label{15a}
\mathrm{(P4):}\max_{\mathbf{Q}, \mathbf{P}, \mathbf{A}, \mathbf{T}}~\sum_{i=1}^{M}u_i,
\end{equation}
\begin{equation}\label{15b}
\text{s.t.}~R_i\geq u_i,\forall i,
\end{equation}
\begin{equation}\label{15c}
~u_i\geq r^*,\forall i,
\end{equation}
\begin{equation}\label{15d}
~P_i\geq 0,\forall i,
\end{equation}
\begin{equation}\label{15e}
~\sum_{i=1}^{M}P_i\leq P_\mathrm{max},
\end{equation}
\begin{equation}\label{15f}
~\alpha_{ij}\in\{0,1\}, \forall i,j,
\end{equation}
\begin{equation}\label{15g}
~\alpha_{ii}=0, \forall i,
\end{equation}
\begin{equation}\label{15h}
~\alpha_{ij}+\alpha_{ji}=1, \forall i\neq j,
\end{equation}
\begin{equation}\label{15i}
~\alpha_{ij}(H^2+\|\mathbf{Q}-\mathbf{q}_i\|^2)\leq H^2+\|\mathbf{Q}-\mathbf{q}_j\|^2, \forall i\neq j,
\end{equation}
\end{subequations}
where $\mathbf{U}=\{u_i,\forall i\}$ are auxiliary variables and all the other notations are the same as those in problem (P2) that remove the time variable $n$. However, problem (P4) is still non-convex due to the non-convex constraints (\ref{15b}), (\ref{15i}) and the binary constraint (\ref{15f}). In the following, we focus on dealing with these non-convex constraints.

\subsection{Deployment Position and Power Control Optimization}
Note that constraint (\ref{15f}) can be presented equivalently as
\begin{subequations}\label{eq16}
\begin{equation}\label{16a}
0\leq \alpha_{ij}\leq 1,\forall i,j,
\end{equation}
\begin{equation}\label{16b}
\alpha_{ij}-\alpha_{ij}^2\leq 0,\forall i,j.
\end{equation}
\end{subequations}
Constraint (\ref{16a}) is affine now; however, constraint (\ref{16b}) is still non-convex due to the convexity of the term $\alpha_{ij}^2$. Based on SCA technique, we can replace the term $\alpha_{ij}^2$ with its first order Taylor expansion at the given feasible point $\bar{\alpha}_{ij}$. However, if we directly apply SCA technique to (\ref{16b}), there will be some iterations in which the problem is infeasible \cite{7841981}. Follow \cite{7841981,8482444}, we introduce nonnegative auxiliary variables $\{\varphi_{ij},\forall i,j\}$, the penalty parameter $\lambda$ and then rewrite the objective function and constraint (\ref{16b}) as
\begin{equation}\label{eq17}
\sum_{i=1}^{M}u_i-\lambda\sum_{i=1}^{M}\sum_{j=1}^{M}\varphi_{ij},
\end{equation}
\begin{equation}\label{eq18}
\alpha_{ij}-\alpha_{ij}^2\leq \varphi_{ij},\forall i,j,
\end{equation}
respectively. It has been be proven in \cite{7841981} that $\varphi_{ij}=0,\forall i,j$ at the convergence points, i.e., (\ref{eq17}), (\ref{eq18}) are equivalent to (\ref{15a}), (\ref{16b}), respectively. Therefore, $\alpha_{ij},\forall i,j,$ can be guaranteed to be 0 or 1 when converging. Now, we can apply SCA technique to (\ref{eq18}) and transform it into
\begin{equation}\label{eq19}
\bar{\alpha}_{ij}^2+2\bar{\alpha}_{ij}(\alpha_{ij}-\bar{\alpha}_{ij})+\varphi_{ij}\geq \alpha_{ij},\forall i,j,
\end{equation}
which is convex. So far, we have transformed the binary constraint (\ref{15f}) into the convex constraints (\ref{16a}) and (\ref{eq19}).

For non-convex constraint (\ref{15i}), it can be rewritten as
\begin{equation}\label{eq20}
\begin{aligned}
\frac{(H^2+\|\mathbf{Q}-\mathbf{q}_i\|^2+\alpha_{ij})^2}{4}\leq H^2+\|\mathbf{Q}-\mathbf{q}_j\|^2\\
+\frac{(H^2+\|\mathbf{Q}-\mathbf{q}_i\|^2-\alpha_{ij})^2}{4},\forall i\neq j.
\end{aligned}
\end{equation}
Note that the left-hand-side (LHS) of (\ref{eq20}) is convex with respect to $\mathbf{Q}$ and $\alpha_{ij}$, and the right-hand-side (RHS) of (\ref{eq20}) is convex with respect to $\|\mathbf{Q}-\mathbf{q}_i\|^2$, $\|\mathbf{Q}-\mathbf{q}_j\|^2$ and $\alpha_{ij}$. Similar to (\ref{eq19}), we employ the SCA technique to the RHS of (\ref{eq20}), and then constraint (\ref{eq20}) can be reformulated as constraint (\ref{eq21}) at the top of the next page, where $\bar{\alpha}_{ij}$ and $\bar{\mathbf{Q}}$ are given feasible points. However, constraint (\ref{eq21}) is still non-convex due to the convexity of the terms $\|\mathbf{Q}-\mathbf{q}_j\|^2$ and $\|\mathbf{Q}-\mathbf{q}_i\|^2$ in the LHS. Proceed to employ SCA technique to the LHS of (\ref{eq21}), and then constraint (\ref{eq21}) can be transformed into convex constraint (\ref{eq22}) at the second top of the next page (Here, we assume that $H\gg 1$ to avoid collisions). Until now, constraint (\ref{15i}) has been rewritten as the convex constraint (\ref{eq22}).

\newcounter{mytempeqncnt1}
\begin{figure*}[ht]
\normalsize
\setcounter{mytempeqncnt1}{\value{equation}}
\setcounter{equation}{20}
\begin{equation}\label{eq21}
\begin{aligned}
\frac{(H^2+\|\mathbf{\bar{Q}}-\mathbf{q}_i\|^2-\bar{\alpha}_{ij})(H^2+\|\mathbf{Q}-\mathbf{q}_i\|^2-\alpha_{ij})}{2}+H^2+\|\mathbf{Q}-\mathbf{q}_j\|^2\geq\frac{(H^2+\|\mathbf{Q}-\mathbf{q}_i\|^2+\alpha_{ij})^2}{4}\\
+\frac{(H^2+\|\mathbf{\bar{Q}}-\mathbf{q}_i\|^2-\bar{\alpha}_{ij})^2}{4}, \forall i\neq j.
\end{aligned}
\end{equation}
\setcounter{equation}{21}
\end{figure*}

\newcounter{mytempeqncnt}
\begin{figure*}[ht]
\normalsize
\setcounter{mytempeqncnt}{\value{equation}}
\setcounter{equation}{21}
\begin{equation}\label{eq22}
\begin{aligned}
\frac{(H^2+\|\mathbf{\bar{Q}}-\mathbf{q}_i\|^2-\bar{\alpha}_{ij})(H^2+\|\mathbf{\bar{Q}}-\mathbf{q}_i\|^2+2(\mathbf{\bar{Q}}-\mathbf{q}_i)^T(\mathbf{Q}-\mathbf{\bar{Q}})-\alpha_{ij})}{2}\!+\!H^2\!+\!\|\mathbf{\bar{Q}}-\mathbf{q}_j\|^2\!+\!2(\mathbf{\bar{Q}}-\mathbf{q}_j)^T(\mathbf{Q}-\mathbf{\bar{Q}})\\
\geq\frac{(H^2+\|\mathbf{Q}-\mathbf{q}_i\|^2+\alpha_{ij})^2}{4}+\frac{(H^2+\|\mathbf{\bar{Q}}-\mathbf{q}_i\|^2-\bar{\alpha}_{ij})^2}{4}, \forall i\neq j.
\end{aligned}
\end{equation}
\hrulefill
\setcounter{equation}{22}
\end{figure*}

For non-convex constraint (\ref{15b}), we first rewrite it as
\begin{equation}\label{eq23}
\mathrm{log}_2\left(1+\frac{\frac{\gamma_0P_i}{H^2+\|\mathbf{Q}-\mathbf{q}_i\|^2}}{1+\sum_{j=1,j\neq i}^{M}\frac{\gamma_0\alpha_{ij}Pj}{H^2+\|\mathbf{Q}-\mathbf{q}_j\|^2}}\right)\geq u_i,\forall i.
\end{equation}
Then we introduce auxiliary variables $\{z_i,\forall i\}$, $\{v_i,\forall i\}$ and reformulate (\ref{eq23}) as
\begin{subequations}\label{eq24}
\begin{equation}\label{24a}
\log_2(1+e^{z_i-v_i})\geq u_i,\forall i,
\end{equation}
\begin{equation}\label{24b}
\frac{\gamma_0P_i}{H^2+\|\mathbf{Q}-\mathbf{q}_i\|^2}\geq e^{z_i},\forall i,
\end{equation}
\begin{equation}\label{24c}
1+\sum_{j=1,j\neq i}^{M}\frac{\gamma_0\alpha_{ij}Pj}{H^2+\|\mathbf{Q}-\mathbf{q}_j\|^2}\leq e^{v_i},\forall i.
\end{equation}
\end{subequations}
Note that constraints (\ref{24a}), (\ref{24b}) and (\ref{24c}) are still non-convex. In the following, we mainly apply SCA technique to transform them into convex constraints. Constraint (\ref{24a}) is non-convex because the LHS is convex with respect to $z_i$ and $v_i$ (the detailed proof is provided in Appendix A). Similar to (\ref{eq21}), constraint (\ref{24a}) can be reformulated as convex constraint
\begin{equation}\label{eq25}
\log_2(1+e^{\bar{z}_i-\bar{v}_i})+\frac{e^{\bar{z}_i-\bar{v}_i}}{1+e^{\bar{z}_i-\bar{v}_i}}(z_i-\bar{z}_i-v_i+\bar{v}_i)\geq u_i,\forall i,
\end{equation}
where $\bar{z}_i$ and $\bar{v}_i$ are given feasible points. For constraint (\ref{24b}), it is equivalent to
\begin{equation}\label{eq26}
\frac{H^2}{\gamma_0P_i}+\frac{\|\mathbf{Q}-\mathbf{q}_i\|^2}{\gamma_0P_i}\leq e^{-z_i},\forall i.
\end{equation}
Note that the LHS of (\ref{eq26}) are convex with respect to $P_i$ and $\mathbf{Q}$ (the detailed proof is provided in Appendix B). However, the RHS of (\ref{eq26}) is also convex with respect to $z_i$, which results in the non-convexity of (\ref{eq26}). Similar to (\ref{eq21}), constraint (\ref{eq26}) can be reformulated as
\begin{equation}\label{eq27}
\begin{aligned}
\frac{H^2}{\gamma_0P_i}+\frac{\|\mathbf{Q}-\mathbf{q}_i\|^2}{\gamma_0P_i}\leq e^{-\bar{z}_i}(1-z_i+\bar{z}_i),\forall i,
\end{aligned}
\end{equation}
which is convex now. For constraint (\ref{24c}), we introduce relaxed variables $\{s_i,\forall i\}$, $\{y_{ij},\forall i,j\}$ and then reformulate it as
\begin{subequations}\label{eq28}
\begin{equation}\label{28a}
s_i\leq H^2+\|\mathbf{Q}-\mathbf{q}_i\|^2 ,\forall i,
\end{equation}
\begin{equation}\label{28b}
\frac{\gamma_0\alpha_{ij}P_j}{s_j}\leq y_{ij},\forall i\neq j,
\end{equation}
\begin{equation}\label{28c}
1+\sum_{j=1,j\neq i}^{M}y_{ij}\leq e^{v_i},\forall i.
\end{equation}
\end{subequations}
However, constraints (\ref{28a}), (\ref{28b}) and (\ref{28c}) are still non-convex. Next, we concentrate on transforming them into convex constraints. Note that constraint (\ref{28b}) is equivalent to
\begin{equation}\label{eq29}
\gamma_0(\alpha_{ij}+P_j)^2+(y_{ij}-s_j)^2\leq \gamma_0(\alpha_{ij}-P_j)^2+(y_{ij}+s_j)^2, \forall i\neq j.
\end{equation}
So far, constraints (\ref{28a}), (\ref{28c}) and (\ref{eq29}) are non-convex due to the same reason, i.e., the RHS is convex. Similar to (\ref{eq21}), based on SCA technique, constraints (\ref{28a}), (\ref{28c}) and (\ref{eq29}) can be approximated as convex constraints
\begin{equation}\label{eq30}
s_i\leq H^2+\|\mathbf{\bar{Q}}-\mathbf{q}_i\|^2+2(\mathbf{\bar{Q}}-\mathbf{q}_i)^T(\mathbf{Q}-\mathbf{\bar{Q}}),\forall i,
\end{equation}
\begin{equation}\label{eq31}
1+\sum_{j=1,j\neq i}^{M}y_{ij}\leq e^{\bar{v}_i}(v_i-\bar{v}_i+1),\forall i,
\end{equation}
\begin{equation}\label{eq32}
\begin{aligned}
\gamma_0(\alpha_{ij}&\!+\!P_j)^2\!+\!(y_{ij}\!-\!s_j)^2\leq 2(\bar{y}_{ij}\!+\!\bar{s}_j)(y_{ij}\!+\!s_j)\!-\!(\bar{y}_{ij}\!+\!\bar{s}_j)^2\\
&\!+\!\gamma_0[2(\bar{\alpha}_{ij}\!-\!\bar{P}_j)(\alpha_{ij}\!-\!P_j)\!-\!(\bar{\alpha}_{ij}\!-\!\bar{P}_j)^2],\forall i\neq j,
\end{aligned}
\end{equation}
respectively, where $\bar{P}_j$, $\bar{s}_j$ and $\bar{y}_{ij}$ are given feasible points. Until now, constraint (\ref{15b}) has been transformed into a convex form, which is approximated as (\ref{eq25}), (\ref{eq27}), (\ref{eq30}), (\ref{eq31}) and (\ref{eq32}).

According to the above discussions, the problem (P1) is approximated as the following problem (P5).
\begin{subequations}\label{eq33}
\begin{equation}\label{33a}
\mathrm{(P5):}\max_{\substack{\mathbf{\Psi},\mathbf{Q},\mathbf{A},\mathbf{P},\\ \mathbf{S},\mathbf{U},\mathbf{V},\mathbf{Y},\mathbf{Z}}}\sum_{i=1}^{M}u_i-\lambda\sum_{i,j}\varphi_{ij},
\end{equation}
\begin{equation}\label{33b}
\begin{aligned}
\text{s.t.}~(\ref{15c}),(\ref{15d}),(\ref{15e}),(\ref{15g}),\\
(\ref{15h}),(\ref{16a}),(\ref{eq19}),(\ref{eq22}),\\
(\ref{eq25}),(\ref{eq27}),(\ref{eq30}),(\ref{eq31}),(\ref{eq32}),
\end{aligned}
\end{equation}
\begin{equation}\label{33c}
\varphi_{ij}\geq 0,\forall i,j,
\end{equation}
\end{subequations}
where $\mathbf{\Psi}=\{\varphi_{ij}, \forall i,j\}$, $\mathbf{S}=\{s_i, \forall i\}$, $\mathbf{V}=\{v_i, \forall i\}$, $\mathbf{Y}=\{y_{ij}, \forall i,j\}$ and $\mathbf{Z}=\{z_i, \forall i\}$. In the next subsection, we will propose an algorithm to solve problem (P5).

\subsection{Overall Algorithm}

Baesd on the above description, problem (P5) is now a convex optimization problem, which can be efficiently solved by convex optimization solvers, such as CVX. The proposed iterative algorithm for problem (P5) is concluded in Algorithm \ref{AL1}. In the inner loop, we update the UAV deployment position and the power control until convergence. In the outer loop, we initially set the penalty parameter $\lambda$ as a sufficiently small value to provide enough degree of freedom for $\alpha_{ij}$, and then we update the penalty parameter $\lambda$ step by step to make sure $\varphi_{ij}\rightarrow0$ at the convergence points. The initialization for Algorithm 1 will be discussed in Section \uppercase\expandafter{\romannumeral4}. Algorithm \ref{AL1} has the complexity of $I_1I_2\mathcal{O}(M^7)$ \cite{6891348}, where $I_1$, $I_2$ are the numbers of the inner and the outer iterations, respectively.


\section{Initialization and Low Complexity Algorithm}
\label{Initialization for algorithm 1}
The initial points for Algorithm \ref{AL1} should be carefully considered because of the following two reasons. On one hand, if we set the initial points randomly, it is prone to make problem (P4) infeasible, especially when $r^*$ is large. On the other hand, the initial points have a great influence on the rate of convergence, ranging from a few minutes to tens of minutes. To deal with the both troubles, we propose an initialization scheme for Algorithm \ref{AL1} in this section. We first derive the analytical solution to power control $\mathbf{P}$ for given UAV deployment position $\mathbf{Q}$. Then we determine the other initial points based on $\mathbf{P}$. Finally, thanks to the characteristic of the users' rate obtained by $\mathbf{P}$, a low complexity algorithm is proposed.

\begin{algorithm}[t]
\caption{Iterative algorithm for problem (P4)}
\label{AL1}
\begin{algorithmic}[1]
\STATE Initialize \{$\mathbf{Q}^0$, $\mathbf{A}^0$, $\mathbf{P}^0$, $\mathbf{S}^0$, $\mathbf{V}^0$, $\mathbf{Y}^0$, $\mathbf{Z}^0$\} and $N_0$; Let $r=0$, $\mathrm{num}=0$.
\REPEAT
\REPEAT
\STATE Solve problem (P5) for given \{$\mathbf{Q}^r$, $\mathbf{A}^r$, $\mathbf{P}^r$, $\mathbf{S}^r$, $\mathbf{V}^r$, $\mathbf{Y}^r$, $\mathbf{Z}^r$\} and obtain the optimal solution \{$\mathbf{Q}^{r+1}$, $\mathbf{A}^{r+1}$, $\mathbf{P}^{r+1}$, $\mathbf{S}^{r+1}$, $\mathbf{V}^{r+1}$, $\mathbf{Y}^{r+1}$, $\mathbf{Z}^{r+1}$\}.
\STATE Update $r=r+1$.
\UNTIL{The fractional increase of the objective value is below a threshold {$\varepsilon_1$}.}
\IF{$\max\{\varphi_{ij}\}>\varepsilon_2$}
\STATE $\lambda=c\lambda$.
\ELSE
\STATE $\mathrm{num}=\mathrm{num}+1$.
\ENDIF
\UNTIL{$\mathrm{num}\geq N_0$}
\end{algorithmic}
\end{algorithm}

\subsection{Analytical Solution to Power Control}

When removing the time variable $n$, (\ref{eq11}) can be simplified as
\begin{equation}\label{eq34}
\sum_{i=1}^{M}R_i=\log_2\left(1+\sum_{i=1}^{M}P_i\tilde{h}_i\right).
\end{equation}
Besides, considering the function $\log_2(1+x)$ is a monotonically increasing function. Therefore, for given UAV deployment position $\mathbf{Q}$, problem (P4) is simplified as
\begin{subequations}\label{eq35}
\begin{equation}\label{35a}
\mathrm{(P6):}\max_{\mathbf{P}}~\sum_{i=1}^{M}P_i\tilde{h}_i
\end{equation}
\begin{equation}\label{35b}
\text{s.t.}~(\ref{15d}),(\ref{15e}),
\end{equation}
\begin{equation}\label{35c}
R_i\geq r^*,\forall i.
\end{equation}
\end{subequations}
To solve problem (P6), we first sort the sequence $\mathbf{\tilde{h}}=\{\tilde{h}_1,\tilde{h}_2,\ldots,\tilde{h}_M\}$ as $\mathbf{\tilde{h}}_s=\{\tilde{h}_{(1)},\tilde{h}_{(2)},\ldots,\tilde{h}_{(M)}\}$, which satisfying $\tilde{h}_{(1)}\leq \tilde{h}_{(2)}\leq \ldots\leq \tilde{h}_{(M)}$. Denote $\mathbf{P}_s=\{P_{(i)},\forall i\}$, $\mathbf{R}_s=\{R_{(i)},\forall i\}$ and $\mathbf{A}_s=\{\alpha_{(i)(j)},\forall i,j\}$, where $P_{(i)}$ and $R_{(i)}$ are the transmission power and the achievable rate of the $(i)$-th user, respectively, $\alpha_{(i)(j)}$ is the decoding order determined by $\tilde{h}_{(i)}$ and $\tilde{h}_{(j)}$. Under this condition, $\alpha_{(i)(j)}=1,\forall j<i$, otherwise, $\alpha_{(i)(j)}=0$. Consequently, $R_{(i)}$ is given by
\begin{equation}\label{eq36}
R_{(i)}=\log_2\left(1+\frac{P_{(i)}\tilde{h}_{(i)}}{1+\sum_{j=1}^{i-1}P_{(j)}\tilde{h}_{(j)}}\right),\forall i.
\end{equation}
Then problem (P6) can be reformulated as problem (P7) because (P7) is only a reorganization of (P6).
\begin{subequations}\label{eq37}
\begin{equation}\label{37a}
\mathrm{(P7):}\max_{\mathbf{P}_s}~\sum_{i=1}^{M}P_{(i)}\tilde{h}_{(i)},
\end{equation}
\begin{equation}\label{37b}
\text{s.t.}~\sum_{i=1}^{M}P_{(i)}\leq P_\mathrm{max},
\end{equation}
\begin{equation}\label{37c}
P_{(i)}\geq 0,\forall i,
\end{equation}
\begin{equation}\label{37d}
R_{(i)}\geq r^*,\forall i.
\end{equation}
\end{subequations}

Note that the objective function and constraints (\ref{37b}), (\ref{37c}) are affine, only constraint (\ref{37d}) is non-convex with respect to $\mathbf{P}_s$. Luckily, constraint (\ref{37d}) is equivalent to
\begin{equation}\label{eq38}
P_{(i)}\tilde{h}_{(i)}\geq(2^{r^*}-1)\left(1+\sum_{j=1}^{i-1}P_{(j)}\tilde{h}_{(j)}\right),\forall i,
\end{equation}
which is affine too. As a result, problem (P6) can be equivalently formulated as the following linear programming (LP).
\begin{subequations}\label{eq39}
\begin{equation}\label{39a}
\mathrm{(LP):}\max_{\mathbf{P}_s}~\sum_{i=1}^{M}P_{(i)}\tilde{h}_{(i)},
\end{equation}
\begin{equation}\label{39b}
\text{s.t.}~(\ref{37b}),(\ref{37c}),(\ref{eq38}).
\end{equation}
\end{subequations}

To solve (LP), the Lagrange function is represented as (\ref{eq40}) at the top of the next page, where $\mu$, $\eta_i$ and $\nu_i,\forall i$ are the Lagrange multipliers. Then the Karush-Kuhn-Tucher (KKT) conditions can be represented as
\newcounter{Lagrange}
\begin{figure*}[htb]
\normalsize
\setcounter{Lagrange}{\value{equation}}
\setcounter{equation}{39}
\begin{equation}\label{eq40}
\begin{aligned}
L(\mathbf{P}_s,\mu,\eta,\nu)=-\sum_{i=1}^{M}P_{(i)}\tilde{h}_{(i)}\!+\!\mu\left(\sum_{i=1}^{M}P_{(i)}\!-\!P_\mathrm{max}\right)\!-\!\sum_{i=1}^{M}\eta_iP_{(i)}\!+\!\sum_{i=1}^{M}\nu_i\left[(2^{r^*}\!-\!1)\left(1\!+\!\!\sum_{j=1}^{i-1}\!P_{(j)}\tilde{h}_{(j)}\right)\!-\!P_{(i)}\tilde{h}_{(i)}\right].
\end{aligned}
\end{equation}
\hrulefill
\setcounter{equation}{40}
\end{figure*}

\begin{equation}\label{eq41}
(\ref{37b}),(\ref{37c}),(\ref{eq38}),
\end{equation}
\begin{equation}\label{eq42}
\mu\geq 0,\eta_i\geq 0,\nu_i\geq 0,\forall i,
\end{equation}
\begin{equation}\label{eq43}
\eta_iP_{(i)}=0,\forall i,
\end{equation}
\begin{equation}\label{eq44}
\mu\left(\sum_{i=1}^{M}P_{(i)}-P_\mathrm{max}\right)=0,
\end{equation}
\begin{equation}\label{eq45}
\nu_i\left[(2^{r^*}\!-\!1)\left(1\!+\!\!\sum_{j=1}^{i-1}\!P_{(j)}\tilde{h}_{(j)}\right)\!-\!P_{(i)}\tilde{h}_{(i)}\right]=0,\forall i,
\end{equation}
\begin{equation}\label{eq46}
\begin{aligned}
\frac{\partial L}{\partial P_{(i)}}=&-\tilde{h}_{(i)}+\mu-\eta_i-\nu_i\tilde{h}_{(i)}\\
&+(2^{r^*}\!-\!1)\sum_{j=i+1}^{M}\nu_j\tilde{h}_{(j)}=0,\forall i.
\end{aligned}
\end{equation}
Note that $\forall i$, the RHS of (\ref{eq38}) is strictly positive, as $r^*>0$ and $P_{(i)}\geq 0$. So, the LHS of (\ref{eq38}) has to be strictly positive, which implies $P_{(i)}>0$. Furthermore, due to (\ref{eq43}), we arrive at
\begin{equation}\label{eq47}
\eta_i=0,\forall i.
\end{equation}
Substitute (\ref{eq47}) into (\ref{eq46}), we can obtain that
\begin{equation}\label{eq48}
\begin{aligned}
\frac{\partial L}{\partial P_{(i)}}=&-\tilde{h}_{(i)}+\mu-\nu_i\tilde{h}_{(i)}\\
&+(2^{r^*}\!-\!1)\sum_{j=i+1}^{M}\nu_j\tilde{h}_{(j)}=0,\forall i.
\end{aligned}
\end{equation}
For (\ref{eq48}), when $i=M$, we have
\begin{equation}\label{eq49}
\mu=(1+\nu_M)\tilde{h}_{(M)}>0.
\end{equation}
Consequently, based on (\ref{eq44}) and (\ref{eq49}), we can derive that
\begin{equation}\label{eq50}
\sum_{i=1}^{M}P_{(i)}=P_\mathrm{max}.
\end{equation}
For (\ref{eq48}), when $i=M-1$, we have
\begin{equation}\label{eq51}
\begin{aligned}
(1+\nu_{M-1})\tilde{h}_{(M-1)}&=(2^{r^*}\!-\!1)\nu_M\tilde{h}_{(M)}+\mu\\
&=(2^{r^*}-1)\nu_M\tilde{h}_{(M)}+(1+\nu_M)\tilde{h}_{(M)}\\
&>(1+\nu_M)\tilde{h}_{(M)}.
\end{aligned}
\end{equation}
As a result,
\begin{equation}\label{eq52}
\nu_{M-1}>(1+\nu_M)\frac{\tilde{h}_{(M)}}{\tilde{h}_{(M-1)}}-1\overset{(a)}{\geq}0,
\end{equation}
where (a) holds since $\tilde{h}_{(M)}\geq\tilde{h}_{(M-1)}$ and $\nu_M\geq0$. Similarly, we can derive that
\begin{equation}\label{eq53}
\nu_i>0,\forall i\in\mathcal{I}\setminus\{M\}.
\end{equation}
Based on (\ref{eq45}) and (\ref{eq53}), we can know that
\begin{equation}\label{eq54}
(2^{r^*}\!-\!1)\left(1\!+\!\!\sum_{j=1}^{i-1}\!P_{(j)}\tilde{h}_{(j)}\right)\!=\!P_{(i)}\tilde{h}_{(i)},\forall i\in\mathcal{I}\setminus\{M\}.
\end{equation}
According to the above derivation, the optimal power control can be expressed as
\begin{subequations}\label{eq55}
\begin{equation}\label{55a}
\begin{aligned}
P_{(i)}&=\frac{2^{r^*}-1}{\tilde{h}_{(i)}}\left(1+\sum_{j=1}^{i-1}P_{(j)}\tilde{h}_{(j)}\right)\\
&=\frac{2^{r^*}-1}{\tilde{h}_{(i)}}2^{(i-1)r*},\forall i\in\mathcal{I}\setminus\{M\},
\end{aligned}
\end{equation}
\begin{equation}\label{55b}
\begin{aligned}
P_{(M)}&=P_\mathrm{max}-\sum_{i=1}^{M-1}P_{(i)}\\
&=P_\mathrm{max}-\sum_{i=1}^{M-1}\frac{2^{r^*}-1}{\tilde{h}_{(i)}}2^{(i-1)r*}.
\end{aligned}
\end{equation}
\end{subequations}
On the other hand, note that
\begin{equation}\label{eq56}
\begin{aligned}
P_{(M)}&\geq \frac{2^{r^*}-1}{\tilde{h}_{(M)}}\left(1+\sum_{j=1}^{M-1}P_{(j)}\tilde{h}_{(j)}\right)\\
&=\frac{2^{r^*}-1}{\tilde{h}_{(M)}}\left[1+(2^{r^*}-1)\sum_{j=1}^{M-1}2^{(j-1)r^*}\right]\\
&=\frac{2^{r^*}-1}{\tilde{h}_{(M)}}2^{(M-1)r^*},
\end{aligned}
\end{equation}
so (LP) is feasible only when
\begin{equation}\label{eq57}
(2^{r^*}-1)\sum_{i=1}^{M}\frac{2^{(i-1)r^*}}{\tilde{h}_{(i)}}\leq P_\mathrm{max}.
\end{equation}
Now, we can obtain $\mathbf{P}$ based on $\mathbf{P}_s$ and the sort order from $\mathbf{h}$ to $\mathbf{h}_s$.
\subsection{Initialization Scheme}

\begin{algorithm}[t]
\caption{Initialization scheme for Algorithm 1}
\label{AL2}
\begin{algorithmic}[1]
\STATE Set small value of $r_0^*$ , desired $r^*$ and $N_{\mathrm{max}}$. Let $\mathrm{step}=\frac{r^*-r_0^*}{N_{\mathrm{max}}}$.
\FOR{$i=1$ to $M$}
\STATE Deploy the UAV right above the $i$-th user.
\STATE Sort the sequences $\mathbf{h}$ as $\mathbf{h}_s$ and obtain the sort index from $\mathbf{h}$ to $\mathbf{h}_s$.
\STATE Obtain $\mathbf{P}_s^i$ according to (\ref{eq55}).
\STATE Compute $R_{\mathrm{sum}}^i=\log_2(1+\sum_{j=1}^{M}P_{(j)}\tilde{h}_{(j)})$.
\ENDFOR
\STATE $i^*=\mathop{\arg\max}\limits_{1\leq i\leq M}R_{\mathrm{sum}}^i$.
\STATE Set $\mathbf{Q}^0=\mathbf{q}_{i^*}$.
\FOR{$i=1$ to $N_{\mathrm{max}}-1$}
\STATE $r_{\mathrm{temp}}^*=r_0^*+i\cdot\mathrm{step}$.
\STATE Obtain $\mathbf{P}_s$ according to (\ref{eq55}). Initialize $\mathbf{P}^0$ based on $\mathbf{P}_s$ and the sort index from $\mathbf{h}$ to $\mathbf{h}_s$.
\STATE Initialize \{$\mathbf{A}^0$, $\mathbf{S}^0$, $\mathbf{V}^0$, $\mathbf{Y}^0$, $\mathbf{Z}^0$\} when (\ref{eq6}), (\ref{28a}), (\ref{24c}), (\ref{28b}) and (\ref{24b}) hold with equality.
\STATE Solve problem (P5) for $r_{\mathrm{temp}}^*$ via Algorithm 1 and obtain the optimal solution $\mathbf{Q}_{\mathrm{temp}}^*$.
\STATE Update $\mathbf{Q}^0=\mathbf{Q}_{\mathrm{temp}}^*$.
\ENDFOR
\STATE Initialize $\mathbf{P}^0$ for desired $r^*$ according to line 12.
\STATE Initialize \{$\mathbf{A}^0$, $\mathbf{S}^0$, $\mathbf{V}^0$, $\mathbf{Y}^0$, $\mathbf{Z}^0$\} for desired $r^*$ according to line 13.
\end{algorithmic}
\end{algorithm}

According to the results in the above subsection, the analytical solution to the power control can be explained as follows: except the user whose channel gain is the strongest, the transmission power of other users is only used to satisfy the QoS constraint $R_i=r^*$, while the excess power is allocated to the user with the strongest channel gain. Therefore, to maximize the sum rate, we should benefit from the strongest channel.

Understand this principle, we propose an initialization scheme which is summarized in Algorithm \ref{AL2}. First, for small value of $r_0^*$, (\ref{eq57}) is guaranteed to hold, we deploy the UAV right above each user and calculate the corresponding sum rate one by one (lines 2-7). Then we can find the position where the sum rate is the highest (line 8) and deploy the UAV right there (line 9) to accelerate the rate of convergence. Afterwards, instead of setting the initial points for the desired $r^*$ directly, we use $r_{\mathrm{temp}}^*$ to approach $r^*$ and update the initial points step by step (lines 10-16) to guarantee the feasibility of the initial points. Finally, we can obtain the initial points for desired $r^*$ according to the eventual results and lines 17 and 18.

\subsection{Low Complexity Algorithm for Problem (P4)}

\begin{algorithm}[t]
\caption{Low complexity approximation algorithm for problem (P4) when $r^*\leq R^*$}
\label{AL3}
\begin{algorithmic}[1]
\FOR{$i=1$ to $M$}
\STATE Deploy the UAV right above the $i$-th user.
\STATE Sort the sequences $\mathbf{h}$ as $\mathbf{h}_s$ and obtain the sort index $\mathbf{S}_i$ from $\mathbf{h}$ to $\mathbf{h}_s$.
\STATE Obtain $\mathbf{P}_s^i$ according to (55).
\STATE $R_{sum}^i=\log_2(1+\sum_{j=1}^{M}P_{(j)}\tilde{h}_{(j)})$.
\ENDFOR
\STATE $i^*=\mathop{\arg\max}\limits_{1\leq i\leq M}R_{\mathrm{sum}}^i$.
\STATE Deploy the UAV right above $i^*$-th user.
\STATE Obtain $\mathbf{P}$ based on $\mathbf{P}_s^{i^*}$ and $\mathbf{S}_{i^*}$.
\end{algorithmic}
\end{algorithm}

Define $R^*=\mathop{\max}\limits_{1\leq i\leq M}{r_i^*}$, where $r_i^*$ is the unique root of (\ref{eq57}) when $\mathbf{Q}=\mathbf{q}_i$, which can be obtained through the bisection method. Then under the condition that $r^*\leq R^*$, the initialization scheme is unnecessary because there is enough degree of freedom for the UAV deployment position. Therefore, we propose a low complexity approximation algorithm which is summarized in Algorithm \ref{AL3}. The explanation of Algorithm \ref{AL3} is similar to that of Algorithm \ref{AL2}, and it is ignored here resultantly. The complexity of Algorithm \ref{AL3} is dominated by the sort operation. At the worst, the sort operation has the complexity of $\mathcal{O}(M^2)$. Therefore, Algorithm \ref{AL3} has the complexity of $\mathcal{O}(M^3)$, which is far less than that of Algorithm \ref{AL1}. The relation between pages and Algorithm \ref{AL1}-Algorithm \ref{AL3} is summarized in Fig. \ref{tab:1} for easy reading.

\begin{figure}[htb]
\centering
\begin{tabular}{cc}
\toprule
Algorithm & Page  \\
\midrule
Algorithm \ref{AL1}  & 6 \\
Algorithm \ref{AL2}  & 8 \\
Algorithm \ref{AL3}  & 8 \\
\bottomrule
\end{tabular}
\captionsetup{font=footnotesize}
\caption{Relation between pages and Algorithm \ref{AL1}-Algorithm \ref{AL3}}
\label{tab:1}
\end{figure}

\section{Numerical Results}
\label{Numerical Results}
In this section, we present the numerical results to demonstrate the performance improvement of our proposed iterative algorithm via jointly optimizing the UAV deployment position and power control (denoted as the N-JDP scheme), and our proposed low complexity approximation algorithm (denoted as the N-LC scheme). We consider a system with $M=4$ users, who are marked with red pentagrams within an area of size $400\times 400$ $\mathrm{m}^2$ in Fig. \ref{position}. The other parameters are set as: $P_\mathrm{max}=1$ W, $H=100$ m, $\gamma_0=10^6$. Besides, we can work out that $R^* \approx 1.09$ bps/Hz in this case.

For comparison, we consider the following two benchmark schemes:
\begin{itemize}
\item N-FDP: A UAV-enabled uplink NOMA system, where the UAV is deployed as a base station at the geometric center of all users and we only optimize the transmission power.
\item FDMA: A UAV-enabled uplink FDMA system which is similar to OMA-TYPE-I in \cite{7973036}, where the available bandwidth is normalized and each user occupies one $M$-th of the available bandwidth. We jointly optimize the transmission power and the UAV deployment position.
\end{itemize}

\begin{figure}[tp]
  \centering
  \includegraphics[width=0.48\textwidth]{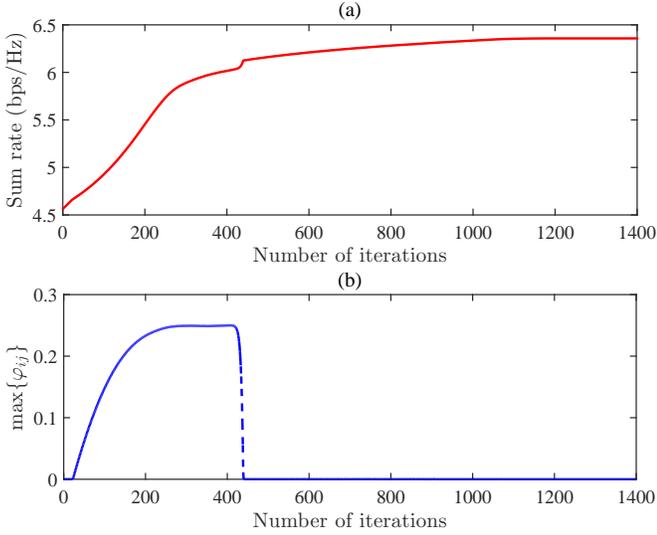}
  \captionsetup{font=footnotesize}
  \caption{Convergence performance of Algorithm 1 with $r^*$=0.5 bps/Hz.}
  \label{convergence1}
\end{figure}

\begin{figure}[tp]
  \centering
  \includegraphics[width=0.48\textwidth]{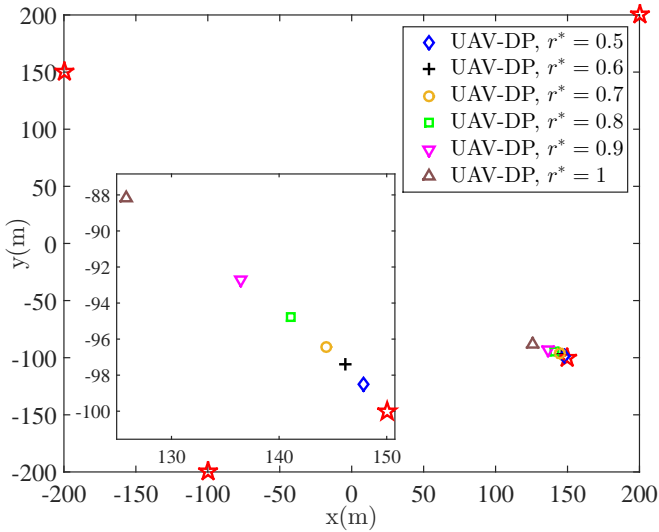}
  \captionsetup{font=footnotesize}
  \caption{Optimal deployment position with N-JDP scheme versus $r^*$.}
  \label{position}
\end{figure}

\begin{figure}[tp]
  \centering
  \includegraphics[width=0.48\textwidth]{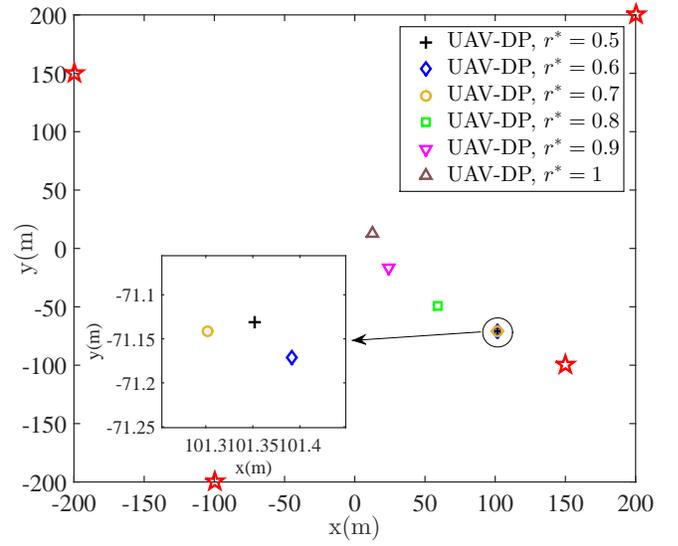}
  \captionsetup{font=footnotesize}
  \caption{Optimal deployment position with FDMA scheme versus $r^*$.}
  \label{position_FDMA}
\end{figure}

\begin{figure}[tp]
  \centering
  \includegraphics[width=0.48\textwidth]{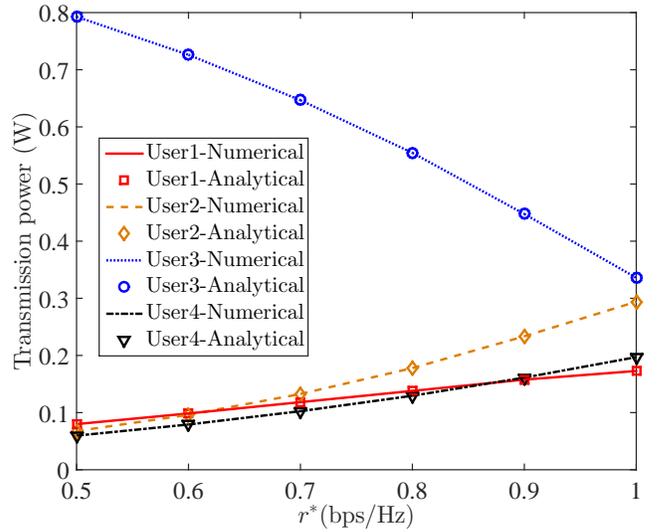}
  \captionsetup{font=footnotesize}
  \caption{Power control with N-JDP scheme versus $r^*$.}
  \label{power}
\end{figure}

In Fig. \ref{convergence1}, we numerically demonstrate the convergence performance of our proposed Algorithm 1. To make it more obvious, we set the initial UAV deployment position as the geometric center of all users. In Fig. 2(a), as we can see, Algorithm 1 converges in the end. Besides, Fig. 2(b) verifies that $\varphi_{ij}\rightarrow0,\forall i,j$ at the convergence points.

Fig. \ref{position} and Fig. \ref{position_FDMA} plot the optimal UAV deployment position (UAV-DP) with our proposed N-JPD scheme and FDMA scheme versus different $r^*$, respectively. For our proposed N-JDP scheme, it is observed that the optimal UAV deployment position is close to user 3. This is because the following reasons: 1) the UAV tries to make use of the strongest channel to maximize the sum rate efficiently and the closer to a certain user, the better the strongest channel. 2) compared with deploying the UAV right above other users, deploying the UAV right above user 3 costs the minimal transmission power for poorer users, which allows allocating the most transmission power to the strongest user. We can also note that the UAV moves away from user 3 as $r^*$ increases. The reason is that other users whose channel gains are poorer need more transmission power to satisfies the QoS constraint as $r^*$ increases. The UAV moving away from user 3 can improve the channel gains of other users to reduce the increase in the transmission power of other users, enabling user 3 to transmit with as much power as possible. This phenomenon can be treated as a tradeoff between the strongest channel gain and the channel gains of other users. The tradeoff can also explain why the optimal UAV deployment position is not the right above of user 3. For the FDMA scheme, however, the difference is that the optimal UAV deployment position quickly approaches the users' geometric center as $r^*$ increases, because there is the least path loss \cite{8663350} and no inter-user interference.

Fig. \ref{power} illustrates the transmission power of ground users with our proposed N-JPD scheme and FDMA scheme versus different $r^*$. It is clearly shown that the numerical results are well matched with the analytical results. Then we can observe that the transmission power of user 3 is the largest, while the transmission power of the other users is small. This is because the channel gain of user 3 is the strongest, increasing its transmission power is the most efficient to improve the sum rate. At the same time, the other users communicate with the UAV at the power that just satisfies $R_i=r^*$, which does not need too much transmission power. We can also observe that as $r^*$ increases, the transmission power of user 3 decreases, while the transmission power of other users increases because they need more transmission power to satisfy the QoS constraint. Consequently, the transmission power of the user 3 whose channel gain is the strongest decreases according to (55b). Furthermore, it is noted that the transmission power of user 1 is the largest among users 1, 2, and 4 when $r^*=0.5$ bps/Hz but the smallest when $r^*=1$ bps/Hz. This can be interpreted that user 1 has the smallest channel gain, and according to (55a), its transmission power is the least sensitive to the increase in $r^*$, but is the most sensitive to the increase in channel gain.

\begin{figure}[tp]
  \centering
  \includegraphics[width=0.48\textwidth]{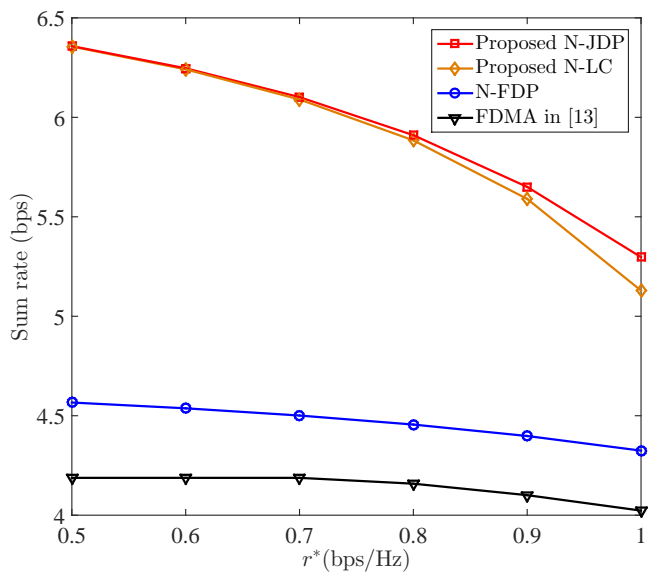}
  \captionsetup{font=footnotesize}
  \caption{Sum rate versus $r^*$.}
  \label{sum-rate}
\end{figure}

\begin{figure}[tp]
  \centering
  \includegraphics[width=0.48\textwidth]{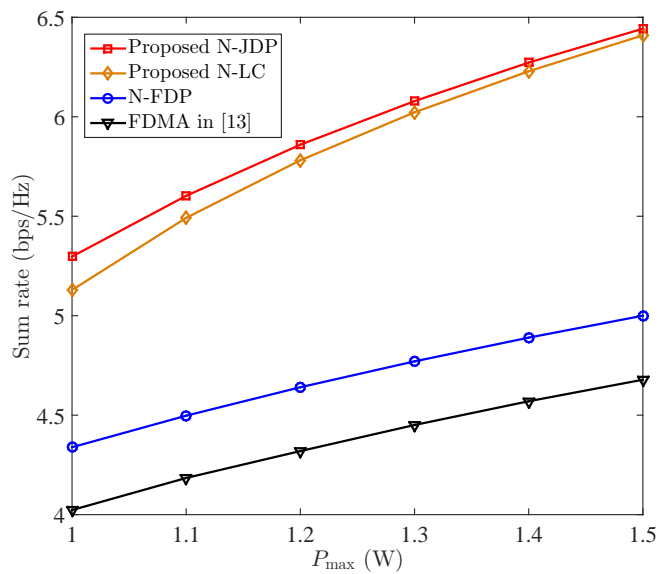}
  \captionsetup{font=footnotesize}
  \caption{Sum rate versus $P_{\mathrm{max}}$ with $r^*=1$ bps/Hz.}
  \label{srpmax}
\end{figure}

\begin{figure}[tp]
  \centering
  \includegraphics[width=0.48\textwidth]{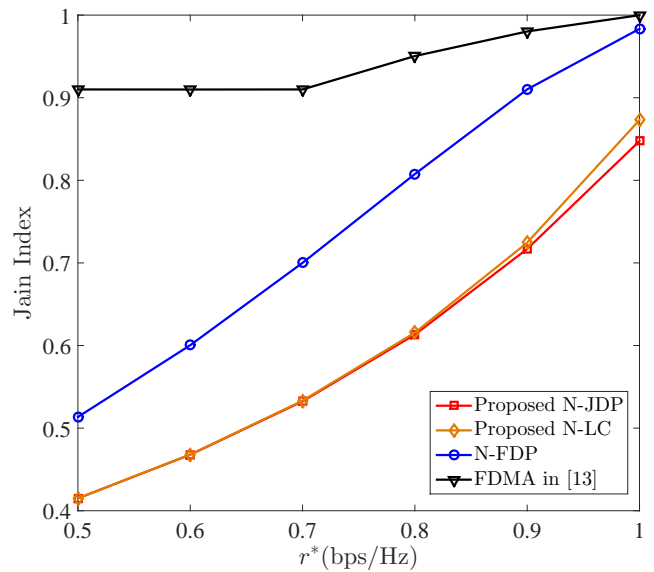}
  \captionsetup{font=footnotesize}
  \caption{Jain index versus $r^*$.}
  \label{fair}
\end{figure}

In Fig. \ref{sum-rate}, we compare the sum rate of the above four schemes versus different $r^*$. For three NOMA schemes, as we expected, the sum rate decreases as $r^*$ increases. This is due to the fact that the larger $r^*$ provides the less degree of freedom for the power control, i.e., it ``wastes" more transmission power on the users whose channel gains are poorer. However, for the FDMA scheme, the power control strategy is quite different due to the disappearance of inter-user interference. Except the minimum power for satisfying the QoS constraint, the excess power is allocated by the water-filling policy \cite{7973036}. As a result, the sum rate remains unchanged when $r^*\leq0.7$ bps/Hz because the optimal rates of all users are greater than $r^*$, while the sum rate decreases when $r^*\geq 0.8$ bps/Hz due to the same reason of NOMA schemes. We can also observe that the sum rate of three NOMA schemes is greater than that of the FDMA scheme. This result indicates that NOMA has higher spectral efficiency than OMA. Besides, the performance of the N-FDP scheme is far less than those of the N-JPD scheme and the N-LC scheme, which demonstrates the performance improvement caused by the UAV deployment position optimization. Last but not least, the performance of our proposed N-LC scheme is very close to that of the N-JPD scheme when $r^*\leq R^*$ and it can even reach more than 96\% of the optimal performance when $r^*=1$ bps/Hz.

Fig. \ref{srpmax} shows the sum rate versus $P_{\mathrm{max}}$ with $r^*=1$ bps/Hz for different schemes. As we expected, for all schemes, the sum rate increases as $P_{\mathrm{max}}$ increases. Then, we can observe that the sum rate of three NOMA schemes outperforms than that of FDMA scheme. Furthermore, it is noted that the performance difference between N-JDP scheme and N-LC scheme decreases as $P_{\mathrm{max}}$ increases because the larger $P_{\mathrm{max}}$, the larger transmission power of user 3 and the UAV consequently tends to be deployed right above user 3. On the contrary, the performance difference between N-JDP scheme and N-FDP scheme increases as $P_{\mathrm{max}}$ increases because the UAV deployment position plays a more and more important role in performance improvement.

In Fig. \ref{fair}, we plot the fairness performance of four schemes. In this paper, we adopt the Jain index as the fairness measurement, which is given by $J=\frac{(\sum_{i=1}^{M}R_i)^2}{M\sum_{i=1}^{M}R_i^2}$. Obviously, the Jain indices of four schemes all increase as $r^*$ increases because the UAV has to communicate with the poorer users at a greater rate. It is interesting to observe that the relationship between the Jain indices of four schemes is completely opposite to the relationship between the sum rates of four schemes. This can be explained that, in NOMA schemes, the UAV tries to enhance the rate of the strongest user as much as possible, but communicates with other users at the minimal required rate. However, in FDMA scheme, it is better to improve the rates of all users at the same time. Therefore, the QoS requirement needs to be carefully considered to strike a balance between the sum rate and the fairness performance.

\section{Conclusions}
\label{Conclusions}
In this paper, we investigated the UAV-enabled uplink NOMA system. To maximize the sum rate of all users, we first proofed that the UAV should stay stationary at a certain point. Then, an iterative algorithm has been proposed to jointly optimize the UAV deployment position and the power control by SCA technique and penalty function method. Due to the high complexity of the iterative algorithm and its initialization, we proposed a low complexity approximation scheme, which has a similar performance but far less complexity compared with the iterative algorithm when $r^*\leq R^*$. Numerical results revealed that our proposed NOMA schemes significantly outperform the conventional NOMA scheme and the OMA scheme such as FDMA.

\appendices
\section{}
Consider the function $f(x,y)=\log_2(1+e^{x-y})$, where $x,y\in\mathbb{R}$. The Hessian matrix of $f(x,y)$ is presented by
\begin{equation}
\begin{aligned}
\nabla^2f(x,y)&=\frac{e^{x-y}}{(1+e^{x-y})^2\ln2}\begin{bmatrix}
1 & -1\\
-1 & 1
\end{bmatrix}\\
&=\frac{e^{x-y}}{(1+e^{x-y})^2\ln2}\begin{bmatrix}
                                                 1 \\
                                                 -1
                                               \end{bmatrix}\begin{bmatrix}
                                                              1 & -1
                                                            \end{bmatrix}\succeq \mathbf{0}.
                                               \end{aligned}
\end{equation}
As a result, $\log_2(1+e^{z_i-v_i})$ is a convex function with respect to $z_i$ and $v_i$.
\section{}
Consider the function $g(\mathbf{x},y)=\frac{\|\mathbf{x}-\mathbf{a}\|^2}{y}$, where $\mathbf{x}\in\mathbb{R}^n$, $y>0$ and $\mathbf{a}\in\mathbb{R}^n$. The Hessian matrix of $g(\mathbf{x},y)$ is presented by
\begin{equation}
\begin{aligned}
\nabla^2g(\mathbf{x},y)&=\frac{2}{y^2}\begin{bmatrix}
                                       y\mathbf{I_n} & -\mathbf{x} \\
                                       -\mathbf{x}^T & \frac{\mathbf{x}^T\mathbf{x}}{y^3}
                                     \end{bmatrix}\\
                                     &=\frac{2}{y^2}\begin{bmatrix}
                                                                  \sqrt{y}\mathbf{I_n} \\
                                                                  \frac{-\mathbf{x}^T}{\sqrt{y}}
                                                                \end{bmatrix}\begin{bmatrix}
                                                                               \sqrt{y}\mathbf{I_n} & \frac{-\mathbf{x}}{\sqrt{y}}
                                                                             \end{bmatrix}\succeq \mathbf{0},
                                     \end{aligned}
\end{equation}
where $\mathbf{I_n}$ is the identity matrix of order $n$. Consequently, the function $\frac{\|\mathbf{Q}-\mathbf{q}_i\|^2}{\gamma_0P_i}$ is a convex function with respect to $\mathbf{Q}$ and $P_i$. Besides, it is well know that the function $\frac{H^2}{\gamma_0P_i}$ is a convex function with respect to $P_i$. Note that a nonnegative weighted sum of convex functions is still convex, we can obtain that the function $\frac{H^2}{\gamma_0P_i}+\frac{\|\mathbf{Q}-\mathbf{q}_i\|^2}{\gamma_0P_i}$ is a convex function with respect to $\mathbf{Q}$ and $P_i$.

\balance
\bibliographystyle{IEEEtran}
\bibliography{cite.bib}
\end{document}